\documentclass[runningheads]{llncs}
\usepackage[T1]{fontenc}
\usepackage{graphicx}
\usepackage{multirow}
\usepackage{booktabs}
\usepackage{color}
\usepackage{amsmath}
\usepackage{hyperref}

\begin{document}
\title{DBpedia-Enriched Company Representation for B2B Lead Recommendation}
\author{Yuyan Qian\inst{1,2} \orcidID{0009-0008-9436-2518}\and
Claude Montacie\inst{1} \orcidID{0000-0003-1936-8975} \and
Milan Stankovic\inst{2} \and
Victoria Eyharabide\inst{1}\orcidID{0000-0002-3775-1495}}
\authorrunning{Y. Qian et al.}
\institute{Sorbonne University, STIH, Paris, France 
\and Leadbay, San Francisco, California}
%\url{http://www.springer.com/gp/computer-science/lncs} }
% \and ABC Institute, Rupert-Karls-University Heidelberg, Heidelberg, Germany\\ 

\maketitle              % typeset the header of the contribution
\begin{abstract}
Selecting which companies to approach is a central challenge in business-to-business (B2B) sales, where decisions are often based on manual research and fragmented information sources. Modern B2B sales platforms centralize company records and use learned company embeddings to support tasks such as recommending and prioritizing potential clients. In this study, we investigate whether enriching these company embeddings with Semantic knowledge from DBpedia improves downstream interaction-prediction performance, within a pipeline that integrates structured company attributes and text embeddings deployed on a real B2B platform. We evaluate the learned embeddings on a downstream interaction prediction task using real user feedback data from the platform. Results show that DBpedia enrichment improves downstream performance, with gains observed on ranking and discrimination metrics.

\end{abstract}

\section{Introduction and Business Context}
%\subsection{A Subsection Sample}

Business-to-business (B2B) sales teams must identify which companies to approach among large candidate sets, where such potential targets are treated as leads in the B2B sales context \cite{Monat2011}.  Lead selection relies on company descriptions, sector information, size, and location,  but many company records remain sparse, noisy, or incomplete. B2B sales platforms have begun to centralize company records and learn compact vector representations to support retrieval and recommendation workflows \cite{gonzalez-floresRelevanceLeadPrioritization2025}. Native company records often include brief descriptions that fail to capture technologies, market segments, or business specializations. Therefore, sparse records make lead prioritization harder, especially for less-documented or long-tail companies. External semantic descriptions can more accurately position such companies by providing context that internal records cannot. Related work \cite{anderssonDatasetManagementPowered2025,bothSolidBasedB2BData2025} has reported industrial benefits for data management and data value chains enabled by knowledge graph technologies. DBpedia provides a practical external source of entity-centered descriptions that can be linked to company text fields and encoded with the same text model as native descriptions. 

This paper investigates whether DBpedia enrichment improves the learned representations of companies in \href{https://wow.leadbay.ai}{Leadbay.ai}, a B2B lead recommendation platform deployed in the French and US markets. Company text fields are linked to DBpedia entities, retrieved descriptions are encoded, and DBpedia-derived features are combined with native company information. Evaluation uses historical sales-user interactions with leads as an offline proxy for recommendation quality in the deployed system.

\section{System Overview and Company Representation}

\begin{figure}[h]
\includegraphics[width=\textwidth]{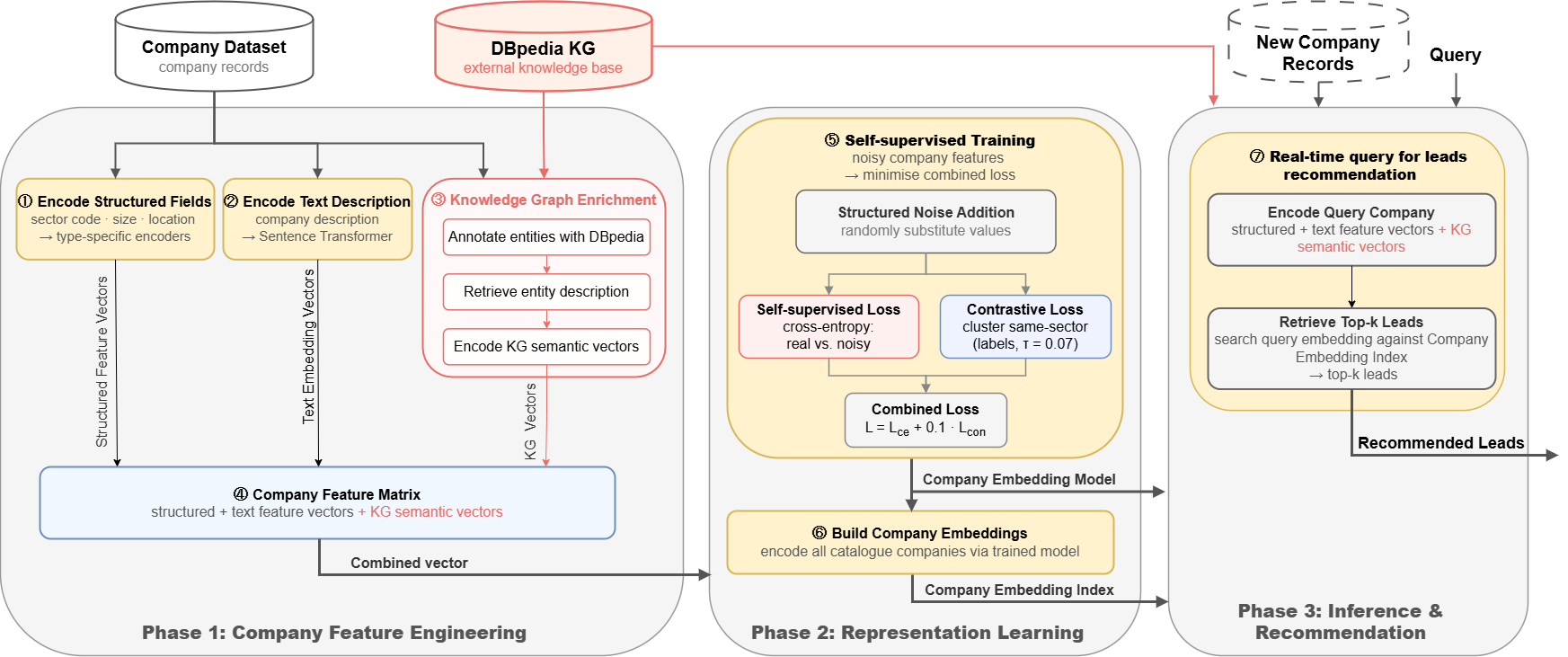}
\caption{Overview of the company representation pipeline, from feature engineering and representation learning to inference-time retrieval of recommended leads.} \label{fig1}
\end{figure}

\href{https://wow.leadbay.ai/}{Leadbay.ai} operates a B2B lead recommendation system built on a centralized repository of companies. A core retrieval module leverages learned representations to encode companies into a shared embedding space. Within this space, candidate leads are ranked according to their similarity to a query company, enabling effective lead prioritization for sales users. The company representation pipeline comprises three phases, as illustrated in Fig.~\ref{fig1}.

%\vspace{-10pt}

Company records are first processed through a feature engineering phase. Structured attributes, such as sector code, company size, and geographic location, are encoded using type-specific encoders, while text descriptions are embedded with \textit{all-MiniLM-L6-v2} \cite{reimers2020makingmonolingualsentenceembeddings}. In addition, company text fields are linked to DBpedia entities, whose corresponding descriptions are encoded using the same model. Native features and DBpedia-derived features are subsequently concatenated to form a unified company feature matrix.    

Company representations are then learned with a TabTransformer-based model \cite{huang2020tabtransformertabulardatamodeling}, trained with a corruption-detection objective and a metadata-informed contrastive regularization scheme based on sector code divisions \cite{chen2024tabdecocomprehensivecontrastiveframework}. 

At inference time, catalogue companies are encoded offline into an embedding index. A new company record is processed through the same pipeline to generate a query embedding. Candidate companies are ranked in the shared embedding space, and the top-$k$ companies are returned as recommended leads.

\section{Experimental Evaluation}
We evaluate the effect of DBpedia enrichment on the quality of company representation using a downstream interaction-prediction task. On the platform, explicit \textit{liked} actions by sales users are treated as positive labels, while all other observed outcomes are treated as non-positive labels. Historical interaction logs serve as an industrial proxy for recommendation quality, since final business outcomes depend on factors beyond the recommendation component. Closed deals depend on sales timing, follow-up quality, account strategy, budget, and other process variables that are not controlled solely by the embedding model. Therefore, interaction logs provide a more direct and earlier signal for representation-level evaluation, as they reflect whether retrieved lead profiles match user interests before later sales stages introduce additional variance.

Better interaction prediction means the retrieval component can rank candidate leads closer to observed user preferences and to lead profiles users judged worth exploring. Higher-quality lead ranking can reduce manual screening effort, encourage more user engagement with surfaced leads, and create better conditions for faster sales follow-up. 

The evaluation compares two settings under enriched and non-enriched conditions: a SentenceTransformer \textit{all-MiniLM-L6-v2} baseline trained only on company descriptions, and a TabTransformer-based representation model trained on structured company attributes and text embeddings. Company representations are evaluated with a logistic regression classifier trained on frozen embeddings. The dataset retains users with at least 15 positive and 15 negative interactions and a positive-to-negative ratio below 10:1, yielding 77 users. Labels are split 80/20 with stratified sampling, and residual class imbalance is handled with \texttt{class\_weight=balanced}. Reported scores are mean values across users.

\vspace{-8pt}
 \begin{table}
 \caption{Interaction prediction performance (mean $\pm$ std over 77 users). Significance vs.\ NoKG counterpart: $^{*}p<0.05$, $^{**}p<0.01$, $^{***}p<0.001$.}
 \label{tab:results}              
 \centering
  \resizebox{\textwidth}{!}{%                  
  \begin{tabular}{lccccc}
  \toprule                            
  \textbf{Model} & \textbf{ROC-AUC} & \textbf{Accuracy} & \textbf{F1} & \textbf{NDCG@10} & \textbf{NDCG@20} \\
  \midrule                                      
  ST-NoKG 
    & $0.719 \pm 0.149$                             
    & $0.710 \pm 0.120$
    & $0.578 \pm 0.205$
    & $0.657 \pm 0.252$
    & $0.699 \pm 0.219$ \\                             
  ST-DBpedia
    & $\mathbf{0.742 \pm 0.141}^{***}$
    & $\mathbf{0.718 \pm 0.129}^{*}$
    & $\mathbf{0.580 \pm 0.208}$
    & $\mathbf{0.692 \pm 0.243}^{*}$
    & $\mathbf{0.721 \pm 0.205}^{*}$ \\                
  \midrule
  TabT-NoKG& $0.654 \pm 0.189$
    & $0.630 \pm 0.147$
    & $0.540 \pm 0.191$
    & $0.597 \pm 0.245$
    & $0.631 \pm 0.223$ \\
  TabT-DBpedia& $0.677 \pm 0.161$
    & $0.659 \pm 0.126^{*}$
    & $0.564 \pm 0.180$
    & $0.615 \pm 0.260$
    & $0.654 \pm 0.227$ \\                                             
  \bottomrule
  \end{tabular}%                               
  }
  \end{table} 
\vspace{-8pt} 

Table~\ref{tab:results} reports the main results. The SentenceTransformer baseline showed significant improvements in ROC-AUC, accuracy, NDCG@10, and NDCG@20 after DBpedia enrichment. The TabTransformer-based model showed smaller gains, with significance only for accuracy. The results indicate that DBpedia-derived semantic descriptions yield the largest benefit when company representations rely primarily on sparse native text, whereas richer structured-plus-text representations yield smaller marginal gains.

\section{Conclusion and Future Work}

DBpedia enrichment yields the largest gains in the SentenceTransformer baseline, which uses only company descriptions. The strongest observed benefit, therefore, appears in data conditions where native representations depend mainly on short text and where internal records provide limited contextual detail. The TabTransformer-based model already combines structured company attributes with text embeddings and achieves smaller gains from DBpedia enrichment. Thus, external Semantic Web knowledge has higher marginal value in weaker or narrower base representations than in richer heterogeneous representations. Adoption also introduces operational cost through entity linking, external knowledge access, and feature augmentation. Entity linking quality, DBpedia coverage, and alignment between DBpedia descriptions and company records therefore affect the return on integration effort. 

In conclusion, we presented an industrial study of DBpedia-enriched company representations for B2B lead recommendation. DBpedia enrichment improved downstream interaction prediction relative to non-enriched counterparts, with the strongest gains observed in the text-based baseline. Future work will examine richer knowledge sources, improved entity linking, additional geographic markets, multilingual company datasets, and online metrics beyond interaction-based offline proxies.

\bibliographystyle{splncs04}
\bibliography{mybibliography}




% LNCS Homepage, \url{http://www.springer.com/lncs}, last accessed 2023/10/25
% \end{thebibliography}
\end{document}